\documentclass[12pt]{iopart}
\usepackage{graphicx}
\usepackage{bm}% bold math
\begin{document}
%-----------------------------------------------
\title{Effective refractive index tensor for weak-field gravity}%
%-----------------------------------------------

\author{%
Petarpa Boonserm \footnote[1]{petarpa.boonserm@mcs.vuw.ac.nz},
Celine Cattoen \footnote[2]{celine.cattoen@mcs.vuw.ac.nz},
Tristan Faber \footnote[3]{tristan.faber@mcs.vuw.ac.nz}, \\
Matt Visser \footnote[4]{matt.visser@mcs.vuw.ac.nz},
and
Silke Weinfurtner \footnote[5]{silke.weinfurtner@mcs.vuw.ac.nz}}
\address{School of Mathematics, Statistics, and Computer Science, \\
Victoria University of Wellington, \\
P.O.Box 600, Wellington, New Zealand}

%-----------------------------------------------

\begin{abstract}

%-----------------------------------------------
  Gravitational lensing in a weak but otherwise arbitrary
  gravitational field can be described in terms of a $3\times3$
  tensor, the ``effective refractive index''. If the sources
  generating the gravitational field all have small internal fluxes,
  stresses, and pressures, then this tensor is automatically isotropic
  and the ``effective refractive index'' is simply a scalar that can
  be determined in terms of a classic result involving the Newtonian
  gravitational potential. In contrast if anisotropic stresses are
  ever important then the gravitational field acts similarly to an
  anisotropic crystal.  We derive simple formulae for the refractive
  index tensor, and indicate some situations in which this will be
  important.

\bigskip

{gr-qc/0411034; 8 November 2004; 

Revised 10 March 2005;  \LaTeX-ed \today}

%-----------------------------------------------
\end{abstract}
%-----------------------------------------------
\pacs{04.20.-q; 04.20.Cv}

%-----------------------------------------------
\maketitle
%-----------------------------------------------

\def\d{{\mathrm{d}}}
%-------------------------------------------------------------------------
\def\ii{{\hat\imath}}
\def\jj{{\hat\jmath}}
\def\kk{{\hat k}}
\def\lll{{\hat l}}
\def\tt{{\hat t}}
\def\xx{{\hat x}}
\def\yy{{\hat y}}
\def\zz{{\hat z}}
%--------------------------------------------------------------------------
\def\eff{{\mathrm{eff}}}

%-----------------------------------------------
\section{Introduction}
%-----------------------------------------------

Weak-field gravity in Einstein's general relativity is actually more
general than straightforward Newtonian gravity~\cite{mtw,schutz}.
While the approximate validity of Newtonian gravity is certainly
limited to the weak-field regime, Newtonian gravity makes significant
additional assumptions as to the smallness of effects that depend on
the internal stresses, pressures, and energy fluxes in the massive
bodies that act as source for the gravitational field. While there is
no significant doubt that for planets, and most stars, the gravitational effects of internal
stresses can safely be neglected, the situation for neutron stars 
(where ${G_N M/ R}\approx {1/10}$) is much more uncertain.
Furthermore, while there is little doubt that the ``dark matter'' that
makes up approximately 90\% of most spiral galaxies can be treated
using weak-field gravity, in the absence of solid physical motivation
for some particular equation of state we cannot necessarily conclude
that the gravitational field can be adequately described by Newtonian
gravity.

In view of this we have developed a formalism that makes no
assumptions about the relative smallness or isotropy of internal stresses (and
pressures and fluxes), to see how gravitational lensing is affected.
In particular, weak Newtonian gravitational lenses can be interpreted
in terms of an analogy wherein a gravitational field is assigned an
``effective refractive index"~\cite{lenses,lenses2}, and we extend
these ideas to see how this ``effective refractive index" is affected
by the presence of significant internal stress. Most strikingly we
will see that the ``effective refractive index" is in general no
longer a scalar, but is instead a $3\times3$ tensor --- in analogy to
the situation in an anisotropic crystal.  (The use of analogies to
relate otherwise distinct phenomena, and to give qualitative insight
as to what physical effects might be important, has recently attracted
significant interest in the general relativity
community~\cite{analogue}, but related ideas under the name
``electro-optical analogy" have an independent history~\cite{nandi}.)
We organize the paper as follows:
\begin{itemize}
  
\item First we consider the static case where there are no internal
  energy fluxes (so in particular we neglect the effects of rotation).
  
\item Second we further specialise this discussion to situations of
  static spherical symmetry --- believed to be a good approximation
  for galactic halos containing ``dark matter''.
  
\item Third we extend the discussion to the more general stationary
  case, where internal fluxes are included.
  
\item Fourth we indicate how time-dependent situations can in
  principle be dealt with.
  
\item Fifth we indicate how the present formalism matches to the usual
  idea of a far-field multipole expansion.
  
\item Finally we briefly discuss astrophysical situations in which the
  issues raised in this article are likely to become important.

\end{itemize}

%-----------------------------------------------
\section{Static case}
%-----------------------------------------------

For light propagating in curved space along some curve parameterized
by $\lambda$ we have
\begin{equation}
g_{ab} \, \d X^a(\lambda) \, \d X^{b}(\lambda) = 0 \, .
\end{equation}
Looking at the specific case of a weak field, where $g_{ab}= \eta_{ab}
+ h_{ab}$, the gravitational field can be considered as a perturbation
$h_{ab}$ around the flat space $\eta_{ab}$. This leads to:
\begin{equation}
g_{ab} \, \d X^a(\lambda) \, \d X^{b}(\lambda) =
\eta_{ab} \, \d X^a(\lambda) \, \d X^{b}(\lambda) +
h_{ab} \, \d X^a(\lambda) \, \d X^{b}(\lambda) = 0 \, .
\end{equation}
Here $\eta_{ab} \, \d X^a(\lambda) \, \d X^{b}(\lambda) $ is no longer
zero.  For a light ray propagating in a \emph{static} weak field we
get
\begin{equation} \label{lightray1}
g_{ab} \, \frac{\d X^a}{\d \lambda} \, \frac{\d X^b}{\d \lambda}  =
(-1+h_{tt})\, \left(\frac{\d t}{\d\lambda}\right)^2 + (\delta_{ij}
+h_{ij})\, \frac{\d x^i}{\d \lambda} \, \frac{\d x^j}{\d \lambda} = 0 \,.
\end{equation}
Choosing $\lambda=t$ equation (\ref{lightray1}) simplifies to
\begin{equation} \label{lightray2}
g_{ab} \, \frac{\d X^a}{\d t} \, \frac{\d X^b}{\d t}  
= (-1+h_{tt}) +
(\delta_{ij} +h_{ij})\, \dot{x}^{i} \, \dot{x}^{j}= 0 \, .
\end{equation}
We define a ``coordinate speed of light'' by calculating the norm
$||\dot x^i ||$ of the ``coordinate velocity of light '' $\dot x^i$
using $\delta_{ij}$, the unperturbed background metric for space. This
allows us to split the velocity into a speed and a direction
\begin{equation}
\dot x^i = || \dot x^i || \; \hat k^i;  \qquad || \hat k^i|| 
= 1 = \sqrt{ \delta_{ij} \;  \hat k^i\; \hat k^j};
\end{equation}
where $\hat k^i$ is a unit 3-vector.  Putting this into equation
(\ref{lightray2}), and noting that the $h_{ab}$ are small compared to
unity we can usefully Taylor series expand, to obtain
\begin{equation}
 || \dot x^i || =  \sqrt{\frac{1-h_{tt}}{1+h_{ij}\, \hat{k}^i \, \hat{k}^j}} 
\approx
1-\frac{1}{2} \, h_{tt} - \frac{1}{2} \, h_{ij} \, \hat{k}^{i} \, 
\hat{k}^{j} + \mathcal{O}(h_{ab}^2)
\end{equation}
for the coordinate speed of light.  Note that we have adopted units
where the physical speed of light, $c$, measured by physical rulers
and physical clocks, is always $1$.  Then the spacetime refractive
index for light travelling in the direction $\hat k$ is
\begin{equation}
n(\hat k) = \frac{1}{ || \dot x^i ||} \approx
1 + \frac{1}{2} \, h_{tt} +  \frac{1}{2} \, h_{ij} \, 
\hat{k}^{i} \, \hat{k}^{j} + \, \mathcal{O}(h_{ab}^2).
\end{equation}
We now define the $3\times3$  \emph{refractive index tensor} as:
\begin{equation}
\label{E:3tensor}
n_{ij} \equiv \left(1 + \frac{1}{2} \, h_{tt}\right) \, \delta_{ij} +
\frac{1}{2} \, h_{ij} \,,
\end{equation}
so that $n(\hat k) = n_{ij} \; \hat k^i\; \hat k^j+  \mathcal{O}(h_{ab}^2)$.
If we adopt the standard definition
\begin{equation}
\bar h^{ab} =  h^{ab} - {1\over2} \; h \; \eta^{ab},
\end{equation}
then a brief computation yields
\begin{equation}
\label{E:3tensor2}
n_{ij} \equiv \left(1 + \frac{1}{2} \, \bar h_{tt}\right) \, \delta_{ij} +
\frac{1}{2} \, \bar h_{ij} \,.
\end{equation}
To connect these general formulae to the presence of stress-energy, adopt
Einstein--Fock--de~Donder gauge (that is, harmonic
quasi-Cartesian coordinates)~\cite{mtw}
\begin{equation}
\label{E:harmonic}
\partial_a \bar h^{ab} = 0,
\end{equation}
and write the Einstein equations in the \emph{exact} form~\cite{mtw}
\begin{equation}
\nabla^2 \bar h_{ab} = -16\pi \; G_N  \; T^\mathrm{eff}_{ab},
\end{equation}
where the ``effective'' stress-energy contains contributions both from ``ordinary'' stress-energy and the ``pseudo-energy'' of the gravitational field itself:
\begin{equation}
 T^\mathrm{eff}_{ab} =  T_{ab} +  t_{ab}.
 \end{equation}
 In all situations we are interested in the pseudo-energy is much smaller than the ordinary stress energy, nevertheless explicitly keeping the pseudo-energy as part of the  effective stress energy is a very useful bookkeeping device. (This is an application of the ``principle of controlled ignorance'' espoused in~\cite{mtw}.)
Due to the assumption
that the spacetime is static, the effective stress-energy tensor is
\begin{equation}
  T^\eff_{ab}=\left[%
\begin{array}{cc}
  \rho^\eff & 0 \\
  0 & T^\eff_{ij} \\
\end{array}%
\right].
\end{equation}
We define
\begin{eqnarray}
\label{E:Phi}
  \nabla^2\Phi &\equiv& 4\pi \;G_N \;\rho^\eff, \\
  \nabla^2\Psi_{ij} &\equiv& 4\pi \;G_N \;T^\eff_{ij},
\end{eqnarray}
where $\Phi$ is the ordinary Newton potential, but with the effective
mass-energy-density $\rho^\eff$ as a source. The $\Psi_{ij}$ are new
post-Newton gravitational potentials arising from the effective internal
pressures and stresses of the source matter.  
Using the Einstein equations we find:
\begin{eqnarray}
%------------
\nabla^2 \bar h_{ij} &=&  -16\pi \;G_N \;T^\eff_{ij}.
\\
%------------
\nabla^2 \bar h_{tt} 
&=&  -16\pi \;G_N\;  \rho^\eff.
 \end{eqnarray}
Imposing suitable boundary conditions at spatial infinity and working
in terms of the density and pressure potentials, we obtain
\begin{eqnarray}
\label{E:wfm}
  \bar h_{tt} &=& -4 \Phi;\\
  \bar h_{ij} &=& -4\Psi_{ij},
\label{E:wfm2}
\end{eqnarray}
and the equivalent (though more complicated) 
\begin{eqnarray}
\label{E:wfm3}
  h_{tt} &=& -2( \Phi +  \delta^{kl} \; \Psi_{kl});\\
  h_{ij} &=& -2 ( 2\Psi_{ij} + \delta_{ij}[\Phi -
  \delta^{kl}\; \Psi_{kl}]).
\label{E:wfm4}
\end{eqnarray}
Thus, the refractive index tensor defined in (\ref{E:3tensor},\ref{E:3tensor2}) takes the
particularly simple form
\begin{eqnarray}
n_{ij}  
\label{ref_ind_tensor_static}
&=& (1-2\Phi)\; \delta_{ij} - 2\Psi_{ij} \, .
\end{eqnarray}
In view of our use of the effective stress-energy, these statements are now all exact, but formal. Because $h_{ab}$ and its derivatives occur in the pseudo-tensor $t_{ab}$, the RHS of these formulae depend implicitly on $h_{ab}$. In practical calculations one would start by setting $h_{ab}\to0$ on the RHS, and then iterate the equations to obtain the desired level of accuracy.

For instance, if the effective internal stresses are isotropic $T^\eff_{ij}\to p^\eff \;
\delta_{ij}$ and $\Psi_{ij} \to \Psi_0 \; \delta_{ij}$, we have the
simplification
\begin{eqnarray}
 h_{tt} &=& -2 (\Phi + 3\Psi_{0});
 \\
 h_{ij} &=& -2 \;\delta_{ij}[\Phi -
\Psi_{0}];
 \\
n_{ij} 
  \label{ref_ind_tensor_static2}
  &=& (1-2\Phi-2\Psi_0)\; \delta_{ij},
\end{eqnarray}
with
\begin{eqnarray}
\label{E:phi}
  \nabla^2\Phi &=& 4\pi \;G_N \;\rho^\eff, \\
\label{E:psi0}
  \nabla^2\Psi_0 &=& 4\pi \;G_N \;p^\eff .
\end{eqnarray}
Thus for isotropic effective stress, the refractive index is a scalar.  If furthermore the gravitational field is weak and smooth enough that $t_{ab} \ll T_{ab}$ this can be turned into an approximate statement about physical perfect fluids, where $T_{ij}\to p\; \delta_{ij}$ --- the refractive index would now be approximately isotropic. (This is the usual situation in most stars: the stellar material is to a good approximation a perfect fluid and the  gravitational pseudo-energy is a small faction of the total mass budget.)

Finally, if the internal stresses are completely negligible
(which is the usual situation in most asteroids, but not stars and planets~\cite{mtw}) we can naively set
$\Psi_{ij}\to 0$ to obtain the Newtonian limit and recover the
well-known standard results~\cite{lenses,lenses2}:
\begin{eqnarray}
  h_{tt} &=& -2 \Phi; \qquad
  h_{ij} = -2 \Phi \;\delta_{ij}; \qquad 
  n_{ij} =   (1-2\Phi)\;\delta_{ij}\, .
\end{eqnarray}
The novelty in the current analysis lies exactly in the manner in
which internal stresses in the body generating the gravitational field
lead to a ``stress potential'' $\Psi_{ij}$ which then influences both
the weak-field metric (\ref{E:wfm}--\ref{E:wfm4}) and the effective
refractive index tensor~(\ref{ref_ind_tensor_static}).

%-------------------------------------------
\section{Static spherically symmetric weak field}
%-------------------------------------------

These general considerations can be made more explicit by working in
situations of spherical symmetry.  Consider now a general static
spherically symmetric weak field, for which the stress energy tensor
(with the background Minkowski metric $\eta_{ab}$ written in terms of
spherical polar coordinates) takes the form
\begin{equation}
T^\eff_{ab} = \left[%
\begin{array}{cccc}
  \rho^\eff(r) & 0 &0&0\\
  0 & p^\eff_r(r) &0&0 \\
  0 & 0 & p^\eff_t(r)\, r^2 & 0\\
  0 & 0 & 0 & p^\eff_t(r)\, r^2 \sin^2\theta
\end{array}%
\right].
\end{equation}
Here $p^\eff_r$ and $p^\eff_t$ are the radial and transverse pressures respectively.
At the origin we must have $p^\eff_r(0) = p^\eff_t(0)$. Because of the way the Einstein equations interact with the harmonic gauge condition, $T^{ab}_\eff$ is exactly conserved in the flat space  covariant sense:
\begin{equation}
[T^{ab}_\eff]_{:b} = 0; 
\qquad \hbox{implying} \qquad 
T^{ab}{}_{:b} = - t^{ab}{}_{:b}
\end{equation}
where the colon denotes the flat-space covariant derivative corresponding to $\eta_{ab}$ in spherical polar coordinates.
Consequently
\begin{equation}
\partial_r p^\eff_r(r) + {2\{p^\eff_r(r)-p^\eff_t(r)\}\over r} =0,
\end{equation}
which we can use to eliminate $p_t(r)$ as
\begin{equation}
p^\eff_t(r) = p^\eff_r(r) + {1\over2}\; r\; \partial_r p^\eff_r(r).
\end{equation}
We can similarly write
\begin{equation}
\bar h_{ab} = \left[%
\begin{array}{cccc}
 H_{0}(r) & 0 &0&0\\
  0 & H_r(r) &0&0 \\
  0 & 0 & H_t(r) r^2 & 0\\
  0 & 0 & 0 & H_t(r) r^2 \sin^2\theta
\end{array}%
\right],
\end{equation}
where at the origin symmetry demands $H_r(0)=H_t(0)$, and the harmonic gauge condition (\ref{E:harmonic}), now in the flat-space spherical polar sense $\bar h^{ab}{}_{:b}=0$,  implies
\begin{equation}
H_t(r) = H_r(r) + {1\over2}\; r\; \partial_r H_r(r).
\end{equation}
After a brief computation
\begin{equation}
\nabla^2 \bar h_{ab} = \bar h_{ab:c}{}^{:c} = \left[%
\begin{array}{cccc}
  Z_{0}(r) & 0 &0&0\\
  0 & Z_r(r)&0&0 \\
  0 & 0 &Z_t(r) \, r^2 & 0\\
  0 & 0 & 0 &  Z_t(r)\, r^2 \sin^2\theta
\end{array}%
\right],
\end{equation}
where we are again using the covariant derivative for flat-space spherical
polar coordinates, and we have defined
\begin{equation}
Z_0 = {1\over r^2} \partial_r \left(r^2 \partial_r  H_0 \right);
\end{equation}
\begin{equation}
Z_r = {1\over r^2} \partial_r \left(r^2 \partial_r  H_r \right)
 - 4 \; { H_r(r)- H_t(r)\over r^2}=  
 {1\over r^4} \partial_r \left(r^4 \partial_r H_r \right);
\end{equation}
\begin{equation}
Z_t = {1\over r^2} \partial_r \left(r^2 \partial_r  H_t \right)
+ 2 \; { H_r(r)-H_t(r)\over r^2} = Z_r + {1\over2}\;r\; \partial_r Z_r.
\end{equation}
The slightly unusual terms proportional to the difference $H_r-H_t$
arise due to the fact that we are now using spherical polar
coordinates (not strictly harmonic coordinates), and because $h_{ab}$
has tensor indices in its own right. The Einstein equations
now yield
\begin{equation}
Z_0 = - 16\pi \;G_N \;\rho^\eff;
\end{equation}
\begin{equation}
Z_r = -16\pi \;G_N \;p_r^\eff;
\end{equation}
plus the redundant equation
\begin{equation}
Z_t = - 16\pi \; G_N \;p_t^\eff.
\end{equation}
These have the formal solutions
\begin{equation}
H_0 = -16\pi\; G_N \; \int \left\{\int \rho^\eff r^2 \d r\right\} r^{-2} \d r = - 4 \Phi;
\end{equation}
\begin{equation}
H_r =  -16\pi\; G_N\; \int \left\{\int p_r^\eff r^4 \d r\right\}  r^{-4} \d r;
\end{equation}
where we have again used equation (\ref{E:Phi}), 
and so
\begin{equation}
H_r - H_t = {8\pi\ G_N\over r^3 } \left\{\int p_r^\eff r^4 \d r\right\}.
\label{E:aniso}
\end{equation}
We can further rearrange this result slightly by defining
\begin{equation}
H_\Sigma \equiv H_r + 2H_t; \qquad H_\Delta \equiv H_r-H_t,
\end{equation}
so that
\begin{equation}
H_r = {H_\Sigma+2H_\Delta\over 3}; \qquad 
H_t = {H_\Sigma-H_\Delta\over 3}.
\end{equation}
For a null curve located at radius $r$, and making an angle $\chi$
with respect to the $\hat r$ direction, equation (\ref{E:3tensor2}) implies
\begin{equation}
n(r,\chi) = 1 +{1\over2} H_0 
+ {1\over2} \left( H_r \cos^2\chi + H_t \sin^2\chi \right),
\end{equation}
so that
\begin{equation}
n(r,\chi) = 1 +{1\over2} H_0 + {1\over 6} H_\Sigma 
+ {1\over6} H_\Delta \;
\left\{3\cos^2\chi -1\right\},
\end{equation}
or
\begin{equation}
n(r,\chi) = 1 -2\Phi + {1\over 6} H_\Sigma 
+ {1\over6} H_\Delta \;
\left\{3\cos^2\chi -1\right\}.
\end{equation}
This particularises equation (\ref{ref_ind_tensor_static}) to the case of spherical symmetry, with  $\Psi_{ij}$ now being written in terms of $H_\Sigma$ and $H_\Delta$, and with explicit integral formulae now being available for these quantities.

Let's compare this with a situation where we know the exact result: If the
object that acts as source for the gravitational field has a definite
surface, with vacuum outside that surface, then application of the
Birkhoff theorem tells us that the spacetime geometry will be
Schwarzschild outside that surface. But because of our subsidiary
assumption, that we are working in the spherical polar version of
Einstein--Fock--de~Donder gauge, there is no remaining freedom in our
coordinate system and we must obtain the weak-field limit of
Schwarzschild geometry in harmonic coordinates --- for which the
effective refractive index is asymptotically isotropic. In harmonic coordinates the Schwarzschild solution has
\begin{equation}
h_{rr} = {2m\over r} - {2m^2\over r^2} + \mathcal{O}(r^{-3});
\end{equation}
\begin{equation}
{h_{\theta\theta}\over r^2}  = {h_{\phi\phi}\over r^2\;\sin^2\theta} = 
  {2m\over r} + {m^2\over r^2};
\end{equation}
\begin{equation}
H_\Delta = - {3m^2\over r^2} + \mathcal{O}(r^{-3}).
\end{equation}
Hence
anisotropies in the effective refractive index are (at least in the
case of spherical symmetry) constrained to rapidly decay outside the source that generates them.
We now show that this is compatible with equation (\ref{E:aniso}) above. Outside the body $T_{ab}=0$ so $p_r^\eff\to t_{rr}$ depends on the pseudo-stress only. But $t_{rr} \sim (\partial h)^2 \sim (m/r^2)^2 = m^2/r^4$. This then implies $H_\Delta \sim m^2 /r^2$, in agreement with the exact result.
The precise distribution of effective refractive index anisotropies 
inside the source object can only be
determined by solving the Einstein equations,
which requires some specific model for the interior distribution of
anisotropic stresses. This is a topic which we hope to explore more
fully in the future.

%-------------------------------------------
\section{Stationary case}
%-------------------------------------------

Let us now consider the stationary case, where
\begin{equation}
  h_{ab} = \left[%
\begin{array}{cc}
  h_{tt} & h_{tj} \\
  h_{it} & h_{ij} \\
\end{array}%
\right],
\end{equation}
and due to symmetry $h_{tj}= h_{jt}$. The condition for a photon
trajectory then becomes:
\begin{equation}
\fl
g_{ab} \, \frac{\d X^a}{\d \lambda} \, \frac{\d X^b}{\d \lambda}  =
(-1+h_{tt})\, \left(\frac{\d t}{\d\lambda}\right)^2 + 2
h_{tj}\,\frac{\d t}{\d\lambda}\,\frac{\d x^j}{\d \lambda} + (\delta_{ij}
+h_{ij})\, \frac{\d x^i}{\d \lambda} \, \frac{\d x^j}{\d \lambda} = 0 \,
.
\end{equation}
Choosing the parameter $\lambda=t$, this becomes a quadratic equation
in $\dot x^i$ with the coordinate speed of light now being given by
\begin{equation}
\fl
||\dot x^i || =
\frac{1}{(\delta_{ij}+h_{ij})\,\hat{k}^i\,\hat{k}^j}
\left(
-\,h_{tj}\,\hat{k}^j + 
\sqrt{  (h_{tj}\,\hat{k}^j)^2 -  
(-1+h_{tt})(\delta_{ij}+h_{ij})\,\hat{k}^i\,\hat{k}^j  }
\right).
\end{equation}
Simplifying, Taylor expanding, and inverting gives the refractive
index
\begin{equation}
  n(\hat{k}) =  1 + \frac{1}{2}\left( h_{tt} + 2 h_{tj}\,\hat{k}^j +
  h_{ij}\,\hat{k}^i\,\hat{k}^j \right) \, + \, \mathcal{O}(h^2),
\end{equation}
which is a very straightforward extension of the static case. But
because of the linear term in $\hat{k}$, it is not possible to bring
this completely into 3-tensor form --- there is additional structure
and we must write
\begin{equation}
n(\hat{k}) = n_{ij}\;\hat{k}^i\,\hat{k}^j \, +h_{tj} \;\hat{k}^j 
+ \, \mathcal{O}(h^2),
\label{E:stat}
\end{equation}
where $n_{ij}$ has exactly the same form as in the static case [see
equation (\ref{E:3tensor})] and the new $h_{tj} \;\hat{k}^j$ term can
be interpreted as being due to motion of the ``effective medium" with
respect to the quasi-Cartesian coordinate system $(t,x^i)$.  This can
be justified by performing a coordinate transformation into the local
rest frame of the ``effective medium'', which is moving with
``velocity" $-h_{tj}$, and exhibits a refractive index tensor
$n_{ij}$.  See \ref{A:moving} for details.  Since the ``medium'' is
generally moving inhomogeneously the local rest frame is not best for
performing explicit calculations --- for practical calculations it is
preferable to fix the coordinate system once and for all and to work
with the $n(\hat k)$ of equation (\ref{E:stat}) above. (In particular,
the differential equations relating the weak field $h_{ab}$ to the
distribution of stress energy are defined in the original coordinate
system, which was chosen to satisfy the Einstein--Fock--de~Donder
gauge condition, and these differential equations do not adapt nicely
to the comoving point of view.)

We can furthermore define additional ``flux potentials''
\begin{equation}
  \nabla^2\Pi_{j} = 4\pi \; G_N \;T_{tj}^\eff
\end{equation}
that couple to the momentum flux.  The corresponding weak-field
Einstein equations are very simple, since $\eta_{ab}$ vanishes for the
off-diagonal elements:
\begin{equation}
 \nabla^2 \bar h_{tj} =  \nabla^2 h_{tj} = -16\pi \; G_N \; T_{tj}^\eff.
\end{equation}
Imposing appropriate boundary conditions
\begin{equation}
  h_{tj} = -4\,\Pi_{j},
\end{equation}
and the refractive index for the stationary case is
\begin{equation}
\label{E:general-stationary}
n(\hat{k}) = n_{ij}(\Phi,\Psi)\;\hat{k}^i\,\hat{k}^j \, 
-4\,\Pi_{j}\,\hat{k}^j + \, \mathcal{O}(\Phi^2,\Psi^2,\Pi^2)
\end{equation}
where  $n_{ij}(\Phi,\Psi)$ is as in equation (\ref{ref_ind_tensor_static}).

An alternative representation is to note that if we define
$k^a=(1;\hat k^i)$ so that $k^a$ is a null vector with respect to the
background metric, then
\begin{equation}
  n(\hat{k}) = 1 + {1\over2} \, h_{ab} \;k^a k^b +  \, \mathcal{O}(h^2)
   = 1 + {1\over2} \, \bar h_{ab} \;k^a k^b +  \, \mathcal{O}(h^2).
\end{equation}
If we further define
\begin{equation}
\Phi_{ab} \equiv 
\left[ \begin{array}{c|c} 
\Phi&\Pi_i\\ \hline \Pi_j& \Psi_{ij} 
\end{array}\right],
\end{equation}
then
\begin{equation}
  n(\hat{k}) = 1 - 2 \,\Phi_{ab}  \; k^a k^b +  \, \mathcal{O}(\Phi_{ab}^2).
\end{equation}

%-----------------------------------------------
\section{Time dependent situations}
%-----------------------------------------------

In the presence of time-dependent sources the only modification is
that the Laplace equations for the potentials should be replaced by
wave equations $\nabla^2 \to -\partial_t^2 + \nabla^2$, which
has the effect of replacing the usual $1/r$ potentials by the
appropriate Li\'enard--Wiechert potential~\cite{censor}. The net
result is that
\begin{equation}
h_{ab}(x,t) = 16\pi G_N \int \d^3y \;
{ \left\{
T^\eff_{ab}(y,\tilde t) - 
{1\over2} \,\eta_{ab} \,T^\eff(y,\tilde t) 
\right\}
\over
|\vec x-\vec y|},
\end{equation}
where $\tilde t$ is the retarded time
\begin{equation}
\tilde t = t - |\vec x-\vec y|.
\end{equation}
But if the so-called null energy condition [NEC] is
satisfied~\cite{censor}, then in particular $T_{ab}\,k^a\,k^b \geq 0$,
implying $h_{ab}\,k^a\,k^b \geq \mathcal{O}(h_{ab}^2)$, so that
$n(\hat k) \geq 1 + \mathcal{O}(h_{ab}^2) $.  That is, the NEC implies
the effective refractive index is [to order $ \mathcal{O}(h_{ab}^2)$]
greater than unity, thereby guaranteeing that in this approximation
the coordinate speed of light is always less than $1$. This connects
the discussion back to the perturbative version of ``superluminal
censorship'' discussed in~\cite{censor}.

%-----------------------------------------------
\section{The far field}
%-----------------------------------------------

For isolated bodies it is possible to expand the far field in terms of
multipole moments. In particular, if we go to the rest frame of the
body, and if the body has time-independent internal structure, then it
is a standard result that~\cite{mtw}
\begin{equation}
h_{ab} = {2 G_N M\over r} \; \delta_{ab} +  \mathcal{O}(1/r^2).
\end{equation}
Thus for a single isolated body
\begin{eqnarray}
n(\hat k) &= 1 +  {1\over2} \,h_{ab} \,k^a k^b +   \mathcal{O}(h^2)
= 1 + {2G_N\, M \over r} +  
\, \mathcal{O}(1/r^2),
\end{eqnarray}
where $k^a=(1;\hat k^i)$ is a null vector with respect to the
background metric. But as long as the total mass is positive $M>0$ we
have $n(\hat k) \geq 1$.  That is, the positive mass theorem implies
that in the far-field regime the effective refractive index is always
greater than unity, thereby guaranteeing that the coordinate speed of
light is always less than $1$. This is a rather different perturbative
version of ``superluminal censorship'', more akin to the ideas
discussed in~\cite{penrose}.  Furthermore note that for an isolated
body at rest the far-field refractive index is automatically isotropic
--- this is consistent with our calculation for situations possessing
spherical symmetry, where we found that anisotropic propagation of
null geodesics was confined to the region inside the body.

If there are many compact objects making up the gravitational lens,
one should simply sum the $1/r$ potentials for each object. If we are
considering a body that is at rest but rotating, then the overall
momentum $P^i$ is zero and the dominant term in the potential $\Pi_i$
comes from the dipole term in the multipole expansion --- this term is
well-known to be related to the total angular momentum of the rotating
body and a standard textbook result yields~\cite{mtw}
\begin{equation}
\Pi_i = 
 {G_N \; (\mathbf{L}\times\mathbf{\hat r})_{\,i} \over 2\,r^2} + \, \mathcal{O}(1/r^3).
\end{equation}
If we consistently retain all $\mathcal{O}(1/r^2)$ terms in a
post-Newtonian analysis then the far field for a stationary rotating
body is (see, for example, equation (19.13) of~\cite{mtw})
\begin{equation}
\fl
h_{ab} = \left[\begin{array}{c|c}
{2 G_N M/ r}- {2 G_N^2 M^2/ r^2} &  
-2 {G_N \; (\mathbf{L}\times\mathbf{\hat r})_{\,j} / r^2}
\\
\hline
-2 {G_N \; (\mathbf{L}\times\mathbf{\hat r})_{\,i} / r^2}
&
({2 G_N M/ r}+ {3 G_N^2 M^2/ 2 r^2}) \delta_{ij}
\end{array}\right] + \, \mathcal{O}(1/r^3)
\end{equation}
which implies
\begin{eqnarray}
n(\hat k) &= 1 + {2G_N\, M \over r} - {G_N^2\, M^2 \over 4 \,r^2}  
- {2 G_N \, (\mathbf{L}\times\mathbf{\hat r}) \cdot \mathbf{\hat k} \over r^2} +
\, \mathcal{O}(1/r^3).
\end{eqnarray}
So angular momentum certainly does leave an imprint on the far-field
refractive index, but in a sub-dominant term.

Of course if the bodies in question are not isolated (one might for
instance be considering a null geodesic passing through the bulk of
the body), then the multipole expansion is not useful --- and one
should resort to the use of the general equation
(\ref{E:general-stationary}) and the linearized field equations for
$\Phi$, $\Pi^i$, and $\Psi^{ij}$ (with $\nabla^2 \to - \partial_t^2 +
\nabla^2$ for time dependent situations).

%-----------------------------------------------
\section{Discussion}
%-----------------------------------------------

Our formulation of weak field [but not necessarily Newtonian]
gravitational lensing is particularly simple and gives a nice
interpretation of the ``effective refractive index'' of the
gravitational field directly in terms of Newtonian-like potentials
coupled to the stress-energy tensor. Doing so results in an
``effective refractive index'' more general than the standard
Newtonian result --- anisotropic stresses are seen to lead to an
anisotropic ``refractive index'' and energy fluxes and angular
momentum are seen to lead to a ``moving medium'' effect.

First we should raise (and settle) the issue of gauge invariance
[coordinate independence] --- our results were obtained by using the
simplifying properties of Einstein--Fock--de~Donder gauge, and so
clearly there is a sense in which the existence and value of the
refractive index tensor depends critically on the use of specific
coordinates. However it must be emphasised that once one has used the
refractive index to calculate quantities such as magnification factors
and/or the angle measured between two images on the sky, these angles
and magnifications have a coordinate-invariant physical meaning
independent of whatever coordinate system was used to carry out the
calculation. So while the analogy that leads to the refractive index
tensor is not itself coordinate invariant, the physical observables
that result at the end of any specific calculation are coordinate
invariants.

We now ask under what circumstances these effects might physically be
important? Are there physical situations in which the gravitational
field is appreciable (but still a weak-field), but with $T_{ij}\approx
T_{tt}$? Perhaps the best known situation of this type arises in the
core of a neutron star where $GM/R\approx 1/10$ so the gravitational
field is reasonably weak, and where matter is approximately described
by a ``stiff'' equation of state, $p=\rho\,c^2$. Furthermore, a
radiation fluid (or neutrino fluid) satisfies $p=\rho\,c^2/3$, so if
radiation fluid ever becomes a significant fragment of the total mass
budget of any clumped system, the ideas of this article would become
important.  The contribution of angular momentum to microlensing,
which in our formalism arises at the dipole level in the $\Pi_i$
potential, has been considered in several articles~\cite{Sereno} using
somewhat different formalism.  A particularly interesting effect is
the possibility of anisotropy in the refractive index --- this occurs
once one moves away from the perfect fluid approximation for
astrophysical bodies, and speculation concerning significant crustal
stresses in neutron stars (and their possible effects on compactness
bounds) is common~\cite{anisotropic}.  Finally we point out that
coherent field configurations, be they electromagnetic fields or
scalar fields, generally induce anisotropic stresses comparable in
magnitude to the energy density. In particular any attempt at
modelling the ``dark matter'' in galactic halos with classical fields
will lead to anisotropic stresses that, while weak, are comparable in
magnitude to the energy density~\cite{dark}.

In summary, so long as there is continuing uncertainty over the total
mass budget and relevant equations of state for the various compact
objects occurring in our universe we feel that it is prudent to retain
the generality of the analysis in the present article.

%-----------------------------------------------

\appendix
%-----------------------------------------------

\section{Moving medium interpretation}

%-----------------------------------------------
\label{A:moving}
%-----------------------------------------------

%
To see how a stationary (non-static) weak gravitational field can be
viewed as a moving ``effective medium", it is useful to temporarily
work with strong gravitational fields.  Pick a particular point in
spacetime $(t_0,x_0^i)$. and consider the change of coordinates
\begin{equation}
t\to \bar t = t; \qquad
x^i \to \bar x^i = x^i - v^i(t_0,x_0) \; [t-t_0]
\end{equation}
with $v^i(t_0,x_0) = g^{ik}(t_0,x_0) \;g_{tk}(t_0,x_0)$.
Then
\begin{equation}
\d \bar t = \d t; \qquad  \d \bar x^i = 
\d x^i + g^{ik}(t_0,x_0) \; g_{tk}(t_0,x_0) \; \d t
\end{equation}
and in the immediate vicinity of $(t_0,x_0^i)$ we have
\begin{eqnarray}
g_{ab} \; \d X^a \; \d X^b &= g_{tt} \; \d t^2 
+ 2 g_{ti}\; \d t\;\d x^i + g_{ij} \; \d x^i \; \d x^j 
\\
&= [g_{tt}- (g_{ti} \;g^{ij} \; g_{jt})] \d\bar t^2 
+ g_{ij} \; \d\bar x^i \; \d\bar x^j
\end{eqnarray}
which clearly has the effect of (locally) banishing the mixed
time-space parts of the metric at the cost that
\begin{eqnarray}
\bar g_{tt} &= g_{tt}- (g_{ti} \;g^{ij} \; g_{jt});
\qquad
\bar g_{ti} = 0;
\qquad
\bar g_{ij} =  g_{ij}.
\end{eqnarray}
But because we wish to apply this in the weak field approximation
$g_{ab} = \eta_{ab} + h_{ab}$ with $h_{ab} \ll 1$, this simplifies
tremendously. Since $g^{ti} = h^{ti} + \mathcal{O}(h^2)$, whereas
$g_{ti}=h_{ti}$ exactly, we see
\begin{eqnarray}
\bar h_{tt} &= h_{tt}+ \mathcal{O}(h^2);
\qquad
\bar h_{ti} = 0;
\qquad
\bar h_{ij} =  h_{ij}.
\end{eqnarray}
That is, in weak field going to the moving coordinates defined by
\begin{eqnarray}
t\to \bar t = t; \qquad
&
x^i \to \bar x^i = 
x^i + \delta^{ik} \;h_{tk}(t_0,x_0) \;[t-t_0] +\mathcal{O}(h^2);
\\
\d \bar t = \d t; \qquad  
&
\d \bar x^i = 
\d x^i + \delta^{ik}\; h_{tk}(t_0,x_0) \; \d t +  \mathcal{O}(h^2);
\end{eqnarray}
allows us (locally) to interpret $-h_{tk}$ as the ``velocity'' of the
medium, while $h_{ij}$ and $h_{tt}$, since they change by at worst
$\mathcal{O}(h^2)$, lead to the same refractive index tensor as was
encountered in the static case.
%

%-----------------------------------------------
\section*{Acknowledgements}
%-----------------------------------------------

This research was supported by the Marsden Fund administered by the
Royal Society of New Zealand.

\clearpage

%-----------------------------------------------
\section*{References}
%-----------------------------------------------

%-----------------------------------------------

%-----------------------------------------------

%-----------------------------------------------
\end{document}